\documentclass[floats,floatfix,showpacs,amssymb,prd,twocolumn,superscriptaddress,nofootinbib]{revtex4-1}

\usepackage{graphicx, epsf, epsfig, amssymb}
\usepackage{bm}
\usepackage{longtable}
\usepackage[usenames,dvipsnames]{xcolor}
\usepackage{color}
\usepackage[breaklinks]{hyperref}
\usepackage{amsfonts,amsmath,amssymb,mathrsfs}

\def\be{\begin{equation}}
\def\ee{\end{equation}}
\def\beq{\begin{eqnarray}}
\def\eeq{\end{eqnarray}}

\begin{document}

\title{Equation-of-state-independent relations in neutron stars}

\author{Andrea Maselli}
\affiliation{Institute of Cosmology $\&$ Gravitation, University of Portsmouth, Dennis Sciama Building,
Portsmouth, PO1 3FX, United Kingdom}

\affiliation{Dipartimento di Fisica, Universit\`a di Roma ``La Sapienza'' \& Sezione, INFN Roma1, P.A. Moro 5, 00185, Roma, Italy.}

\author{Vitor Cardoso}
\affiliation{CENTRA, Departamento de F\'{\i}sica, Instituto Superior
  T\'ecnico, Universidade T\'ecnica de Lisboa - UTL, Avenida~Rovisco Pais
  1, 1049 Lisboa, Portugal}
\affiliation{Perimeter Institute for Theoretical Physics
Waterloo, Ontario N2J 2W9, Canada
}
\affiliation{Department of Physics and Astronomy, The University of Mississippi, University, MS 38677, USA.
}

\author{Valeria Ferrari}
\affiliation{Dipartimento di Fisica, Universit\`a di Roma ``La Sapienza'' \& Sezione, INFN Roma1, P.A. Moro 5, 00185, Roma, Italy.}

\author{Leonardo Gualtieri}
\affiliation{Dipartimento di Fisica, Universit\`a di Roma ``La Sapienza'' \& Sezione, INFN Roma1, P.A. Moro 5, 00185, Roma, Italy.}

\author{Paolo Pani}
\affiliation{CENTRA, Departamento de F\'{\i}sica, Instituto Superior
  T\'ecnico, Universidade T\'ecnica de Lisboa - UTL, Avenida~Rovisco Pais
  1, 1049 Lisboa, Portugal}
\affiliation{Institute for Theory $\&$ Computation, Harvard-Smithsonian
CfA, 60 Garden Street, Cambridge, MA, USA}


\date{\today}

\begin{abstract}
  Neutron stars are extremely relativistic objects which abound in our universe
  and yet are poorly understood, due to the high uncertainty on how matter
  behaves in the extreme conditions which prevail in the stellar core. It has
  recently been pointed out that the moment of inertia $I$, the Love number
  $\lambda$ and the spin-induced quadrupole moment $Q$ of an isolated neutron
  star, are related through functions which are practically independent of the
  equation of state. These surprising universal $I-\lambda-Q$ relations pave the
  way for a better understanding of neutron stars, most notably via
  gravitational-wave emission.
Gravitational-wave observations will probe highly-dynamical binaries and it is
important to understand whether the universality of the $I-\lambda-Q$ relations
survives strong-field and finite-size effects.  We apply a Post-Newtonian-Affine
approach to model tidal deformations in compact binaries and show that the
$I-\lambda$ relation depends on the inspiral frequency, but is insensitive to
the equation of state. We provide a fit for the universal relation, which is
valid up to a gravitational wave frequency of $\sim 900$ Hz and accurate to
within a few percent. Our results strengthen the universality of $I-\lambda-Q$
relations, and are relevant for gravitational-wave observations with advanced
ground-based interferometers. We also discuss the possibility of using the
Love-compactness relation to measure the neutron-star radius with an uncertainty
$\lesssim10\%$ from gravitational-wave observations.
\end{abstract}

\maketitle

\noindent{\bf{\em I. Introduction.}}
Neutron stars (NSs) are extremely compact objects formed as the end-state of the
collapse of massive stars, which populate the universe either in isolation or in
binaries.  NS-NS binaries are one of the most promising sources for
second-generation, ground-based detectors of gravitational waves, such as
Adv. Virgo~\cite{VIRGO}, Adv. LIGO~\cite{LIGO} and KAGRA~\cite{KAGRA} (see also
the proposed third-generation detector ET \cite{ET}); in addition, NSs are
copious radio and X-ray emitters, and they can potentially be used as
laboratories for high-energy and fundamental physics, to probe the behaviour of
matter in extreme conditions~\cite{Lyne:2004cj}.  A persistent obstacle against
exploring the full potential of NSs physics lies precisely in our ignorance on
their inner structure, and in particular in the uncertainties of the equation of
state (EoS) of matter at ultranuclear densities. The star radius $R$, mass $M$,
moment of inertia $I$ and deformability, as measured by the tidal Love number
$\lambda$ and spin-induced quadrupole moment $Q$, all depend sensitively on the
EoS. The lack of knowledge on the EoS therefore affects our understanding of the
NS properties and prevents model-independent tests of gravity with these
objects.

Despite the multitude of modern EoS proposed in the literature, leading to
different NS configurations, Yagi and Yunes (hereafter, YY) recently
discovered~\cite{Yagi:2013bca,Yagi:2013awa} some universal relations between the
moment of inertia, the tidal Love number and the spin-induced quadrupole moment,
which are essentially insensitive to the NS EoS.
These tantalizing $I-\lambda-Q$ relations open the interesting possibility of
breaking degeneracies between these parameters and effectively make NSs viable
laboratories for fundamental physics and astrophysics.

The tidal Love number used to derive the $I-\lambda-Q$ relations was computed
assuming a static, spherically symmetric star placed in a time-independent
external quadrupolar tidal field~\cite{Hinderer:2007mb}. Effectively, this means
that these relations were derived for stars in isolation.  However,
gravitational-wave detectors in the next years will observe coalescing compact
binaries, i.e. they will detect gravitational waves emitted by NSs at small
orbital separations, $d/R\lesssim 10$, and high frequencies
$f_{\textnormal{GW}}\geq 40 \,{\rm Hz}$; thus gravitational-wave observations
will probe highly dynamical NSs which strongly interact with their compact
companion (either a black hole or another NS). The question then arises as to
whether the $I-\lambda-Q$ relations are actually useful or even correct in
situations of physical interest.

The purpose of this article is to derive EoS-independent
$I-\lambda$ relations which hold true throughout almost the entire
inspiralling phase of the binary coalescence, and to provide accurate fits describing this relation at each
orbital frequency.

\noindent{\bf{\em II. Evaluation of the tidal Love number.}}
Tidal deformability properties of NSs can be described in terms of a set of
parameters, the {\it Love numbers}
\cite{Flanagan:2007ix,Hinderer:2007mb,Damour:2009vw,Binnington:2009bb,
  Damour:2009wj}, which relate the mass multipole moments of the star to the
(external) tidal field multipole moments. In particular, the dominant
contribution to the stellar deformation is encoded in the electric, $l=2$ Love
number, which we simply call tidal Love number $\lambda$, and is defined by
the relation:
\begin{equation}
Q_{ij}=-\lambda C_{ij} \,,\label{defl}
\end{equation}
where $Q_{ij}$ is the traceless quadrupole moment of the star, and
$C_{ij}=e_{(0)}^\alpha e_{(i)}^\beta e_{(0)}^\gamma e_{(j)}^\delta
R_{\alpha\beta\gamma\delta}$ is the tidal tensor which induces the deformation;
$e_{(\mu)}^\alpha$ is a parallely transported tetrad attached to the deformed
star, and $R_{\alpha\beta\gamma\delta}$ is the Riemann tensor.

 Two approaches are currently used to evaluate the tidal Love number: a
 stationary and a dynamical approach. In the stationary approach
 used by YY \cite{Flanagan:2007ix,Hinderer:2007mb,Damour:2009vw,Binnington:2009bb,
   Damour:2009wj,Read:2008iy,Vines:2011ud}, the compact bodies forming the
 binary system are assumed to be very far apart. Using spacetime perturbation
 theory \cite{1967ApJ...149..591T} to study the $l=2$ stationary perturbations
 of a NS induced by a test tidal field, the quadrupole and tidal tensors are
 evaluated; the Love number is then computed from Eq.~(\ref{defl}). As discussed
 in \cite{Flanagan:2007ix,Maselli:2012zq}, this approach assumes that the
 timescale of the stellar deformation is much smaller than timescales associated
 to the orbital motion, an assumption which becomes less accurate in the last
 stages of coalescence.

 In the dynamical approach \cite{Ferrari:2011as,Maselli:2012zq}, the evolution
 of the tidal deformation of NSs in compact binaries is modeled combining the
 post-Newtonian (PN) description of the two-body metric and of the orbital
 evolution, with an {\it affine description} of the NS as a deformable
 ellipsoid, subject to its self-gravity, to internal pressure forces and to the
 PN tidal field of the companion. The deformed NS is described in terms of five
 dynamical variables: the principal axes of the ellipsoid, and two angles
 describing the orientation of the principal frame; these quantities are
 determined by solving a set of ordinary differential equations in time, coupled
 with the PN equations of motion.  This approach, called Post-Newtonian Affine
 (PNA), allows to compute $Q_{ij}(t)$ and $C_{ij}(t)$ in terms of the dynamical
 variables, so that the tidal Love number can be evaluated during the inspiral.
 To parametrize the dynamical evolution of the system, it is convenient to use
 the orbital frequency $f$, instead of time or radial distance.  The ratio
 between quadrupole and tidal tensors is then a function (the tidal {\it Love
   function}) $\lambda(f)$~\cite{Ferrari:2011as,Maselli:2012zq}, and the Love
 number obtained in the stationary approach corresponds to the zero-frequency
 (i.e. infinite orbital separation) limit of this function.

 The PNA approach also allows to compute the moments of inertia
 $I_i=I\times(a_i/R)^2$, where $i=1,2,3$ indicate the star principal axes ($i=1$
 corresponds to the axis pointing toward the companion), $I$ and $R$ are the
 moment of inertia and the radius of the spherical star, and $a_i$ is the
 $i$-th axis of the deformed, ellipsoidal star. During the inspiral, $I_1$
 increases, while $I_2$ and $I_3$ decrease.

\noindent{\bf{\em III. Results.}}
%
\begin{figure*}[htb]
\begin{center}
\begin{tabular}{l}
\includegraphics[width=8.2cm]{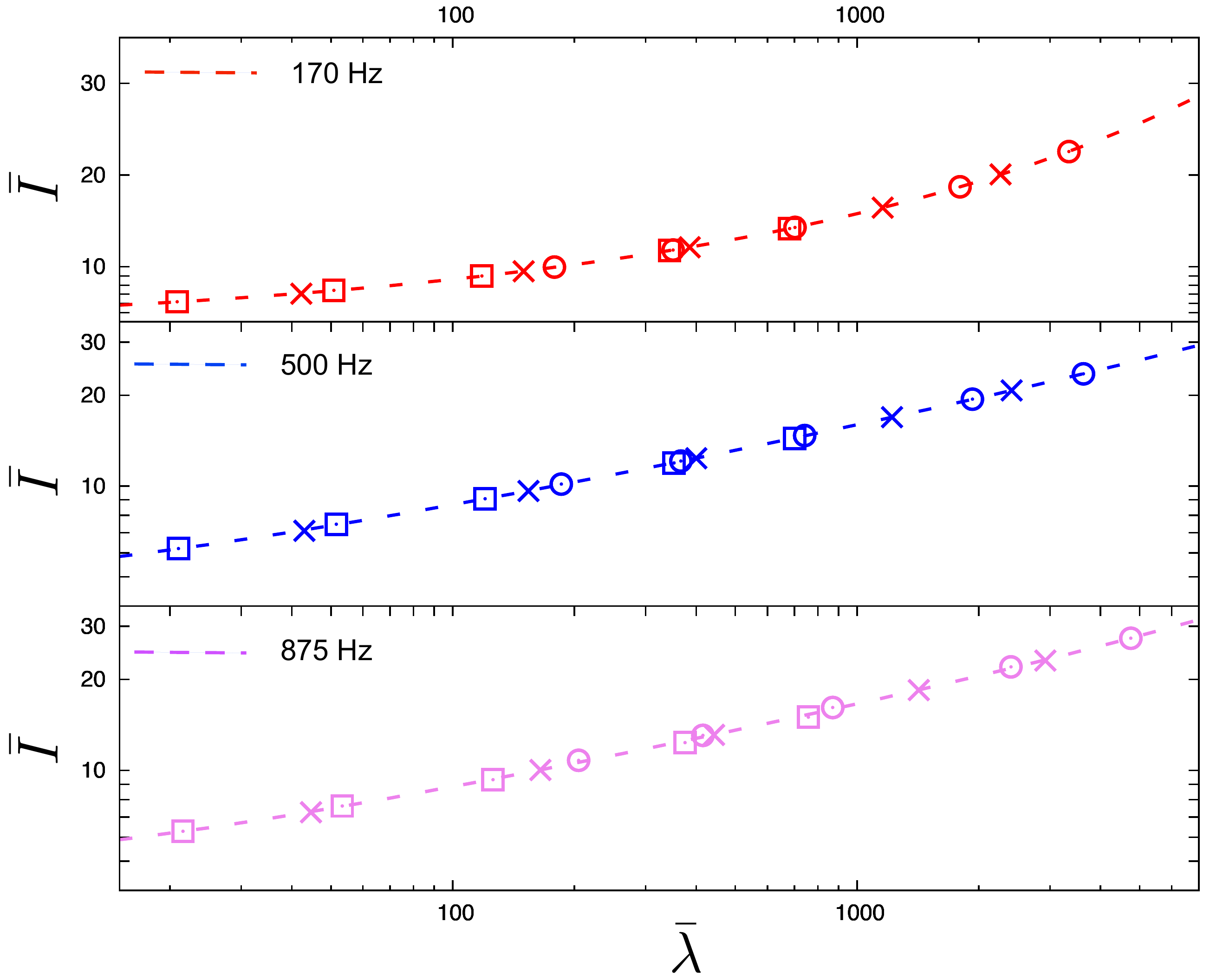}
\hspace{0.5cm} 
\includegraphics[width=8.7cm]{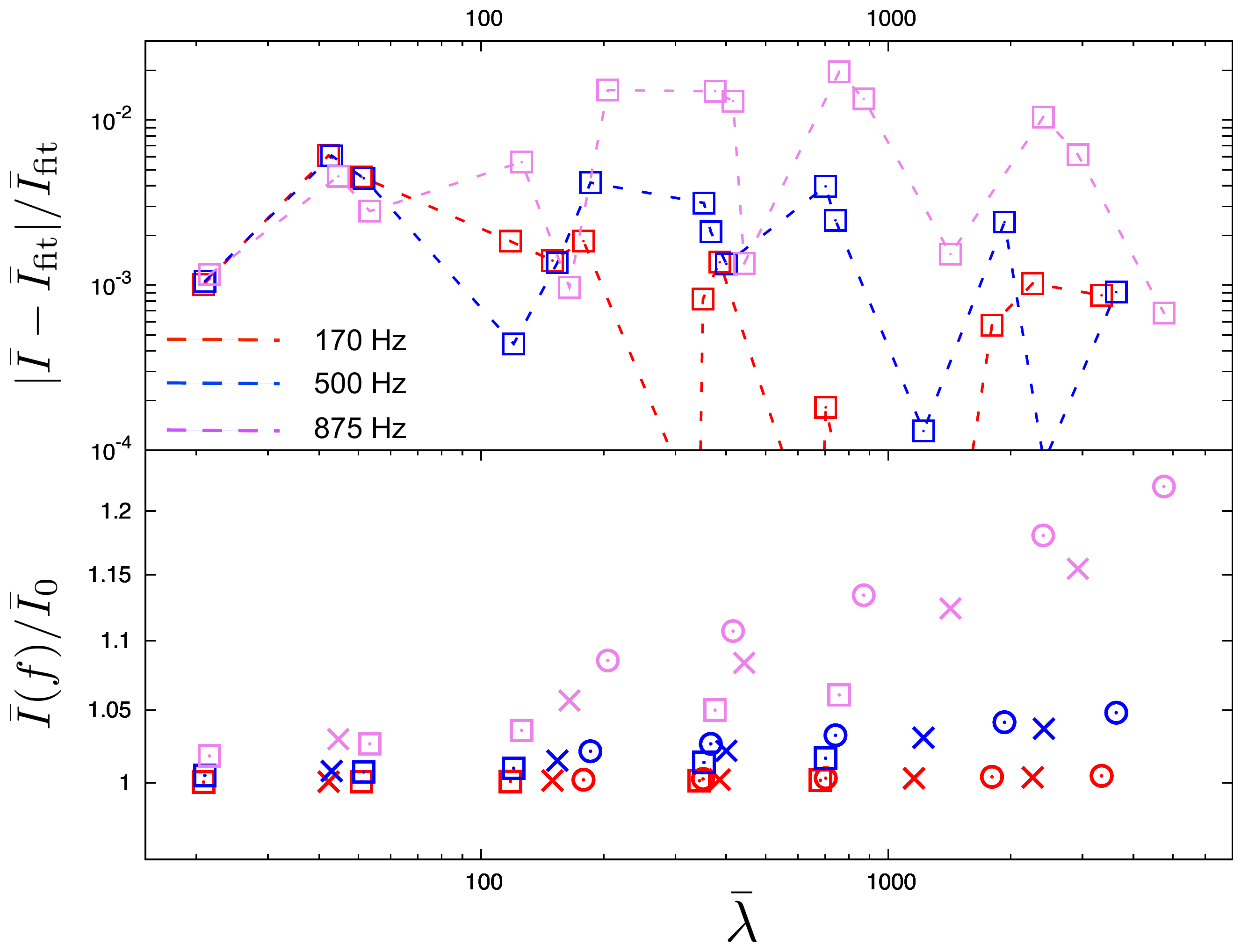}\end{tabular}
\caption{(Left) The $\bar I-\bar \lambda$ relation is plotted for equal 
mass NS-NS binaries, three values of the gravitational wave frequency
$f_{\textnormal{GW}}\equiv 2 f=(170,500,875)$~Hz 
and for the EoS \texttt{APR4} ($\times$), \texttt{MS1} ($\bigcirc$), 
\texttt{H4} ($\Box$). Markers refer to numerical data, while dashed 
lines are the fits~\eqref{fit_gen}.
(Right Top) Relative fractional errors between fits and numerical 
results. (Right Bottom) $\bar I-\bar \lambda$ relation with moment 
of inertia 
normalized by its value at infinity.}
\label{plot_fit}
\end{center}
\end{figure*}
Using the PNA approach, we have computed the normalized Love function
$\bar\lambda=\lambda/M^5$ and the normalized moment of inertia corresponding to
the axis pointing toward the companion, $\bar I=I_1/M^3,$ as functions of the
orbital frequency $f$; $M$ is the NS mass.

We have performed simulations of NS-NS binaries for three different EoS which
are expected to cover a wide range of NS deformability, \texttt{APR4},
\texttt{MS1} and \texttt{H4}, and masses in the range $[1.2\div2]M_{\odot}$ .
In Table~\ref{tableeos} we show the maximum mass, and the radius and compactness
$C=M/R$ of a $1.4M_{\odot}$ star, for the EoS \texttt{APR4}, \texttt{MS1} and
\texttt{H4}. Comparing these values with those shown in the extensive survey
of~\cite{Read:2008iy}, we see that (excluding manifestly unphysical models), all
EoS fall in the range of compactness considered in this paper within $10\%$. The
EoS \texttt{APR4} describes soft NS matter and yields models with high
compactness and small deformability, whereas \texttt{MS1} describes stiff matter
and large deformability; \texttt{H4} provides intermediate configurations. All
EoS are modeled by parametrized piecewise polytropes as proposed by Read
\textit{et al.}~\cite{Read:2008iy}.
\begin{table}[ht]
\centering
\begin{tabular}{ccccc}
\hline
\hline
EOS &  $M_{\textnormal{max}}/M_\odot$ & $R_{1.4}~(km)$ & $C_{\textnormal{1.4}}$  \\  
\hline
$\texttt{APR4}$ & $2.20$ &  $11.12$ & $0.186$\\
$\texttt{H4}$ & $2.03$ &  $13.59$ & $0.152$\\
$\texttt{MS1}$ & $2.78$ & $14.47$ & $0.143$
\end{tabular} 
\caption{Maximum mass,
  radius and compactness of a $1.4M_{\odot}$ neutron star,
  for the EoS \texttt{APR4}, \texttt{H4} and \texttt{MS1}.}
\label{tableeos}
\end {table} 
%
%

Our results are summarized in Fig.~\ref{plot_fit}.  On the three left panels, we
plot $\bar I$ versus $\bar{\lambda}$ for three different values of the
gravitational wave frequency, $f_{\textnormal{GW}}\equiv 2f=170,~ 500,~ 875$ Hz,
for equal-mass NS-NS binaries with different EoS. The data have been fitted with
the following function
\begin{equation}
\ln \bar{I}=b_0+b_1\ln \bar{\lambda}+b_2(\ln 
\bar{\lambda})^{2}+b_3(\ln \bar{\lambda})^{3}+b_4(\ln 
\bar{\lambda})^{4}\,, \label{fit_gen}
\end{equation}
where the fitting parameters $b_i$ are functions of
$f_{\textnormal{GW}}$,
and are listed in Table~\ref{tablefit2}.
\begin{table}[ht]
  \centering
  \begin{tabular}{cccccc}
   \hline
    \hline
    $ f_{\textnormal{GW}}$ & $b_0$ & $b_1$ & $b_2$ & $b_3$ & $b_4$\\  
     \hline
    170 & $1.54$ & $-3.72\cdot 10^{-2}$ & $5.49\cdot 10^{-2}$ & 
$-4.78\cdot10^{-3}$ & $1.87\cdot10^{-4}$\\
    \hline
    300 & $1.58$ & $-6.53\cdot 10^{-2}$ & $6.26\cdot 10^{-2}$ & 
$-5.68\cdot10^{-3}$ & $2.26\cdot10^{-4}$\\
    \hline
    500 & $1.60$ & $-8.34\cdot 10^{-2}$ & $6.83\cdot 10^{-2}$ & 
$-6.39\cdot10^{-3}$ & $2.59\cdot10^{-4}$\\
    \hline
    700 & $1.64$ & $-1.18\cdot 10^{-1}$ & $7.89\cdot 10^{-2}$ & 
$-7.69\cdot10^{-3}$ & $3.18\cdot10^{-4}$\\
    \hline
     800 & $1.68$ & $-1.46\cdot 10^{-1}$ & $8.76\cdot 10^{-2}$ & 
$-8.77\cdot10^{-3}$ & $3.68\cdot10^{-4}$\\
    \hline
    875 & $1.71$ & $-1.72\cdot 10^{-1}$ & $9.54\cdot 10^{-2}$ & 
$-9.72\cdot10^{-3}$ & $4.12\cdot10^{-4}$\\
\hline\hline
    any	& $1.95$  & $-3.73\cdot 10^{-1}$ & $1.55\cdot 10^{-2}$& 
$-1.75\cdot10^{-3}$ & $7.75\cdot10^{-4}$\\
  \end{tabular}
  \caption{Fitting parameters of the $\bar{I}-\bar{\lambda}$ relation 
given by Eq.~\eqref{fit_gen}, for several values of the gravitational
wave frequency.
These fits reproduce our data to within $2\%$, cf. 
Fig.~\ref{plot_fit}. The last row corresponds to the 
fit~\eqref{fit_uni} that reproduces data at \emph{any} frequency to 
within $5\%$ [cf. Fig.~\ref{plot_fit2}].}
 \label{tablefit2}
  \end{table}
%
%
The dashed lines in the left panels of Fig.~\ref{plot_fit} are the fits
corresponding to the selected frequencies.

On the upper, right panel in Fig.~\ref{plot_fit} the relative error
$(\bar{I}-\bar{I}_{fit})/\bar{I}_{fit}$ is plotted versus $\bar{\lambda}$, for
the selected frequencies.  This error is always $\lesssim 2\%$.  In the lower
panel the ratio $\bar{I}(f)/\bar{I}_0$ is plotted versus $\bar{\lambda}$, where
$\bar{I}_0$ is the asymptotic value of $\bar{I}$ when the stars are in
isolation.  This figure shows that, as the stars approach the merger, their
moments of inertia change with respect to the asymptotic value, and grow as much
as $10\%$-$30\%$, depending on the EoS: for stiffer EoS, the variation with
respect to the values at infinity is larger.

Nonetheless, the relative errors
$(\bar{I}-\bar{I}_{fit})/\bar{I}_{fit}$ are small and only mildly dependent on
the EoS, suggesting that a simple frequency-independent relation can be found between $\bar{I}$ and $\bar{\lambda}$.
We find that
\begin{align}
\ln \bar{I}=1.95&-0.373\ln \bar{\lambda}+0.155(\ln 
\bar{\lambda})^{2}\nonumber\\
&-0.0175(\ln \bar{\lambda})^{3}
+0.000775(\ln \bar{\lambda})^{4}\,, \label{fit_uni}
\end{align}
describes very well our numerical results.
%
\begin{figure}[!h]
\centering
\includegraphics[width=8.7cm]{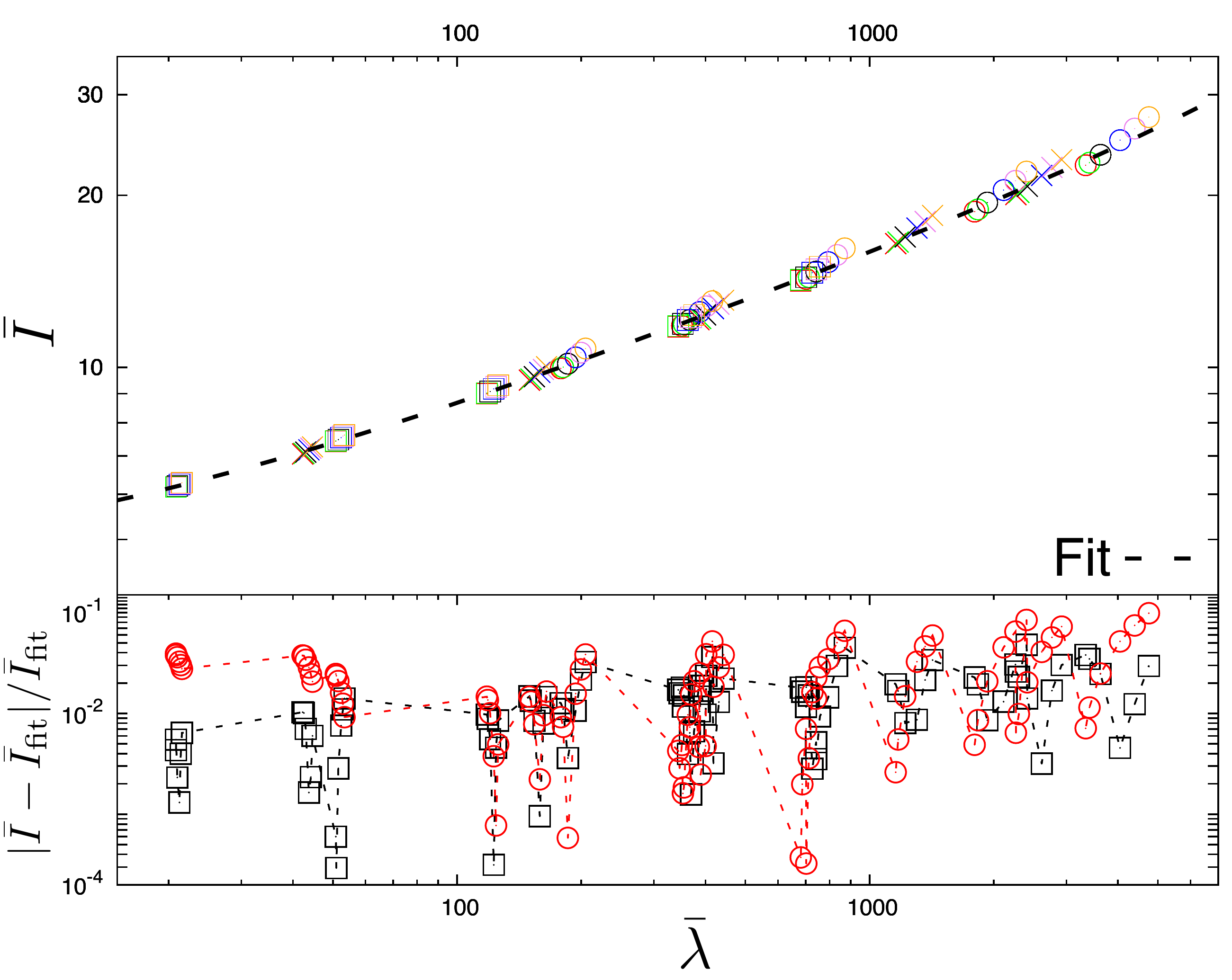}
\caption{(Top) Fitting curve~\eqref{fit_uni} (dashed line) and 
numerical results of the 
$\bar I-\bar \lambda$ relation, for data-set including points up to 
$f_{\textnormal{GW}}=875$ Hz 
and the EOS \texttt{APR4} ($\times$), \texttt{MS1} ($\bigcirc$), 
\texttt{H4} ($\Box$).
(Bottom) Relative fractional errors between fits and numerical 
results.
Black squares and red circles refer to the fit of 
Eq.~(\ref{fit_uni}), 
and to the analytical relation found by~YY, respectively.}
\label{plot_fit2}
\end{figure}
%
In the upper panel of Fig.~\ref{plot_fit2}, we compare the fit \eqref{fit_uni} with our numerical results, 
for the full set of binaries and frequencies we have
considered. In the lower panel we plot the relative errors between the numerical results and the universal fit \eqref{fit_uni},
showing also how well YY's fit performs in the dynamical case.
Our fit (Eq.~\eqref{fit_uni}) reproduces the $\bar I-\bar \lambda$ 
relation to within $5\%$ at any frequency $\lesssim 875$~Hz and for 
all EoS and masses we have considered, while the YY fit 
becomes less accurate as the frequency increases, with fractional 
errors which become of the order of $10\%$.  

The general fit~\eqref{fit_uni} also holds for unequal mass NS-NS
binaries.
For instance, we have checked that for $M_1=1.2\,M_\odot$ and $M_2=1.6\,M_\odot$, the
fit reproduces the $\bar I-\bar \lambda$ relation at any frequency $\lesssim 875$~Hz
to within $5\%$ for the $1.2\,M_\odot$ star, and $3\%$ for the $1.6\,M_\odot$
star.

\noindent{\bf{\em IV. Discussion.}}
NS-NS binaries are the prototypical sources for upcoming second-generation
gravitational-wave detectors.  Strong-field and finite-size effects are
important to model the waveform during the latest stages of the inspiral.  Our
results show that the $I-\lambda$ relations discovered by YY in the low
frequency regime, can be extended to describe the dynamical
evolution of NSs during the late stages of the inspiral.  These results open the
possibility for strong-field, model-independent tests of NS properties. If, for
instance, advanced gravitational wave detectors LIGO/Virgo measure the
tidal Love number in a compact binary coalescence to within $(5-10)\%$, as
estimated in \cite{Flanagan:2007ix,Damour:2012yf}, this would allow for an
indirect estimate of the moment of inertia with roughly the same precision. This
measurement would be independent of (and competitive to) the estimates coming
from pulsar-timing observations~\cite{Lattimer:2004nj}.  In addition, as shown
in YY, these estimates would allow to set constraints on modified theories of
gravity.

Although in this paper we have presented only NS-NS binaries, our approach also
describes the dynamical evolution of mixed black hole-NS systems as well. We
have computed the $I-\lambda$ relation for mixed binaries (with mass-ratio up to
$5$), finding similar universal relations during the entire inspiral.

In this work we have studied the $I-\lambda$ relation, to understand the effects of the tidal
interaction when the stars are at short orbital distance. Spin
effects have been neglected. 
They have been considered by YY in the slow rotation, low frequency limit. 
It would be interesting to establish whether a
simple EoS-independent, universal relation exists, between the tidal Love number
and the spin-induced quadrupole moment $Q$ in the fast rotation, high frequency
regime. This matter will be investigated in a following work.

We conclude this discussion with some considerations on the relation between the
tidal Love number and the NS compactness.  YY showed that this relation is more
EoS-dependent than the $I-\lambda-Q$ relations. However, they included in their
study hot and young NSs, which are unlikely to be members of a coalescing binary
system.  If we consider only old and cold NSs, we find that the $C-\lambda$
relation acquires a remarkable universality \footnote{This is consistent with
  the results of Ref.~\cite{Urbanec:2013fs}, who studied the $Q-C$ relation
  using a set of EoS describing old, cold NSs, finding hints of universality.
  If we combine the $Q-C$ and the $Q-\lambda$ relation discovered by YY, the
  universal behaviour of the $C-\lambda$ relation naturally emerges.}.  By
computing $\bar{\lambda}$ in the low frequency limit, for the EoS \texttt{APR4},
\texttt{MS1}, \texttt{H4}, and masses in the range $[1.2\div2]M_{\odot}$, we
find that $C$ is well described by the fit
\begin{equation}
C=3.71\times10^{-1}\!-3.91\times10^{-2}\ln\bar\lambda
+1.056\times10^{-3}(\ln\bar\lambda)^2\,.
\end{equation}
This fit gives the compactness with a relative error $\lesssim2\%$. 

The $C-\lambda$ relation can be extremely useful to extract information on the
NS EoS from a detected gravitational wave signal emitted in a binary
coalescence. If the tidal Love number is extracted by Advanced LIGO/Virgo with
an error $\sigma_{\ln\lambda}=\sigma_\lambda/\lambda\sim60\%$
\cite{Yagi:2013awa}, we can determine the compactness with an error
$\sigma_C\sim\sqrt{\sigma_{fit}^2+(\partial C/\partial\ln\lambda)^2
  \sigma_{\ln\lambda}^2}\lesssim10\%\,C$ (where we have assumed
$\sigma_{fit}\lesssim{\rm max}(C-C_{fit})$). A much more optimistic estimate of
the error on the measure of the tidal Love number,
$\sigma_{\ln\lambda}\sim5\%$\cite{Damour:2012yf}, would imply a relative error
on the compactness of the order of $\sim2\%$ (this remarkable decrease of the
relative error, can be traced back to the $\sim R^{5}$ dependence of
$\bar\lambda$). Since the same detection would allow for an accurate estimate
of the NS mass, we would then know the NS radius with an uncertainty of $\sim10\%$
or smaller.  It should be noted that current estimates of NS radius based on
astrophysical observations (see \cite{Ozel:2012wu} and references therein), with
a claimed error of $\sim 10\%$, are highly debated in the literature since they
may depend on the way the NS surface emission is modeled \cite{Steiner:2010fz}.
A measurement of the NS radius, based on gravitational wave observations and on
the $C-\lambda$ relation, would have the same, or better, accuracy and it would
be model independent. Such a measurement would be extremely useful to put
constraints on the NS EoS \cite{Lattimer:2000nx}.

\noindent{\bf \em Acknowledgements.}
We would like to thank M. Fortin for useful discussions.  V.C. acknowledges
partial financial support provided under the European Union's FP7 ERC Starting
Grant ``The dynamics of black holes: testing the limits of Einstein's theory''
grant agreement no. DyBHo--256667.  Research at Perimeter Institute is supported
by the Government of Canada through Industry Canada and by the Province of
Ontario through the Ministry of Economic Development and Innovation.
A.M. is supported by a ``Virgo EGO Scientific Forum'' (VESF) grant.
P.P acknowledges financial support provided by the European Community through
the Intra-European Marie Curie contract aStronGR-2011-298297.
This work was supported by the NRHEP 295189 FP7-PEOPLE-2011-IRSES Grant, and by
FCT-Portugal through projects CERN/FP/123593/2011.
%

\bibliographystyle{h-physrev4}

\end{document}